\documentclass[a4paper,fleqn,usenatbib]{mnras}
\usepackage[T1]{fontenc}
\usepackage{ae,aecompl}

\usepackage{graphicx}	
\usepackage{amssymb}	

\usepackage[usenames]{color}
\usepackage{journals}

\newcommand {\comment}[1]{}
\newcommand {\bc}{\begin {centre}}
\newcommand {\ec}{\end {centre}}
\newcommand {\be}{\begin {equation}}
\newcommand {\ee}{\end {equation}}
\newcommand {\beq}{\begin {eqnarray}}
\newcommand {\eeq}{\end {eqnarray}}

\newcommand {\ergs}{erg\,s$^{-1}$}
\renewcommand{\d}{{\rm d}}

\newcommand{\msun}{{\rm M}_{\sun}}

\title[The maximum luminosity for magnetized neutron stars]
{On the maximum accretion luminosity of magnetized neutron stars: 
connecting X-ray pulsars and ultraluminous X-ray sources}
\author[A.~A.~Mushtukov et al.] {Alexander~A.~Mushtukov,$^{1,2}$\thanks{E-mail: al.mushtukov@gmail.com}  
Valery~F.~Suleimanov,$^{3,4}$
Sergey~S.~Tsygankov$^{1}$ 
\newauthor
and  Juri~Poutanen$^{1}$ \\
$^1$Tuorla Observatory, Department of Physics and Astronomy, University of Turku, 
  V\"ais\"al\"antie 20, FI-21500 Piikki\"o, Finland \\
$^2$Pulkovo Observatory of the Russian Academy of Sciences, Saint Petersburg 196140, Russia \\
$^3$Institut f\"ur Astronomie und Astrophysik, Universit\"at T\"ubingen, 
    Sand 1, D-72076 T\"ubingen, Germany \\
$^4$Kazan (Volga region) Federal University, Kremlevskaja str., 18, Kazan 420008, Russia 
}

\begin{document}

\date{Accepted 2015 September 6. Received 2015 August 24; in original form 2015 June 11}

\pubyear{2015}

\pagerange{\pageref{firstpage}--\pageref{lastpage}}

\maketitle
\label{firstpage}

\begin{abstract}
We study properties of luminous X-ray pulsars using a simplified model of the accretion column.
The maximal possible luminosity is calculated as a function of the neutron star (NS) 
magnetic field and spin period.  It is shown that the luminosity can reach values of the order of
$10^{40}\,$\ergs\ for the magnetar-like magnetic field 
($B\gtrsim 10^{14}$~G) and long spin periods ($P\gtrsim 1.5$~s).
The relative narrowness of an area of feasible NS parameters which
are able to provide higher luminosities leads to the conclusion that
$L\simeq 10^{40}\,$\ergs\ is a good estimate for the limiting accretion
luminosity of a NS.  
Because this luminosity coincides with the cut-off  observed in the high mass X-ray binaries luminosity function 
which otherwise does not show any features at lower luminosities, we can conclude that a substantial part of ultra-luminous
X-ray sources are accreting neutron stars in binary systems. 
\end{abstract}


\section{Introduction}
\label{intro}

Accretion of matter on to a compact object in a binary systems is an extremely efficient  way of producing high-energy photons. However, the luminosity that can be reached in this process is limited by different factors. In the case of spherical accretion, the luminosity is restricted by 
the well known Eddington luminosity $L_{\rm Edd}={4\pi GMm_{\rm H}c}/{[\sigma_{\rm T}(1+X)]}\sim 2\times 10^{38}\,$\ergs, 
for a typical mass of the neutron star (NS) $M=1.4\msun$ (here $m_{\rm H}$ is hydrogen mass and $X$ is its mass fraction). 
On the other hand, the luminosity of highly magnetized NS can exceed $L_{\rm Edd}$ substantially for two reasons: 
the interaction cross-section can be much below the Thomson value and 
the radiation can escape in the direction perpendicular to the incoming flow of the material 
which is funnelled to the accretion column \citep{BS1976,LS1988}. The radiation pressure is balanced here by strong magnetic pressure
instead of gravity and the corresponding flux can exceed the Eddington flux significantly.

Some  transient X-ray pulsars indeed produce super-Eddington luminosities  \citep{1976ApJ...206L..25L,1996Natur.379..799K,2006MNRAS.371...19T}. 
Furthermore, a recent revolutionary discovery by the \textit{NuSTAR} observatory 
of pulsations from an ultra-luminous X-ray source (ULX) X-2 in galaxy M82  \citep{Nature2014}  implies that accreting NS 
can reach luminosities of about $10^{40}\,$\ergs, two orders of magnitude above the Eddington limit. 
Until recently one thought that either the intermediate mass black holes $10^2 - 10^5 \msun$ \citep{1999ApJ...519...89C} 
or the super-critically accreting stellar mass black holes \citep[e.g.][]{2007MNRAS.377.1187P}
are the most probable sources of such high X-ray luminosities \citep[see review by][]{2011NewAR..55..166F}.   
Although  highly-magnetized NSs have  previously been suggested to power ULXs  \citep{2013MNRAS.431.2690M}, 
it was rotation rather than accretion that was responsible for high energetics. 

From the standard theory of accretion onto magnetized NSs it is known that the configuration of emitting region highly depends on the mass accretion rate \citep{BS1976}. Particularly, at low luminosity the X-ray emission originates from the hot spots at the NS surface.  As soon as the pulsar luminosity exceeds the critical value ($\sim 10^{37}\,$\ergs) an accretion column begins to rise above the NS \citep{BS1976,2015MNRAS.447.1847M} and provides a possibility for the luminosity to exceed the Eddington limit. 
The radiation pressure is balanced by the magnetic pressure in the accretion column, which supports the column. 
The  height of the accretion column increases with the mass accretion rate and can be as high as the NS radius  \citep{LS1988}. 
Even higher accretion columns are probably possible but their luminosity are not expected to grow significantly because 
of a rapidly decreasing accretion efficiency.  

In this paper, we develop a simplified model of accretion onto magnetized NSs based on the suggestions by  \citet{BS1976} and \citet{LS1988}. 
For the first time, we take into account exact Compton scattering cross-sections in strong magnetic field. 
The structure and the maximal luminosity of the accretion column are calculated.  
We determine an upper limit on the NS luminosity to be close to $10^{40}\,$\ergs, 
which coincides with the position of a sharp cut-off in the luminosity function of high-mass X-ray binaries \citep*{Mineo2012} 
indicating that a large fraction of bright X-ray binaries are likely accreting magnetized NSs. 
A possibility of extremely high mass accretion rate ($\dot{M}\gtrsim 10^{19}\,{\rm g\,s}^{-1}$) onto  a magnetized NS is discussed. 
We find that such high $\dot{M}$ imply extremely high, magnetar-like magnetic field strength.

\section{Accretion column  model}
\label{sec:TheModel}

\subsection{Analytical estimates}
\label{sec:AnalyticalEstimates}

\begin{figure}
\centering 
\includegraphics[width=6.5cm]{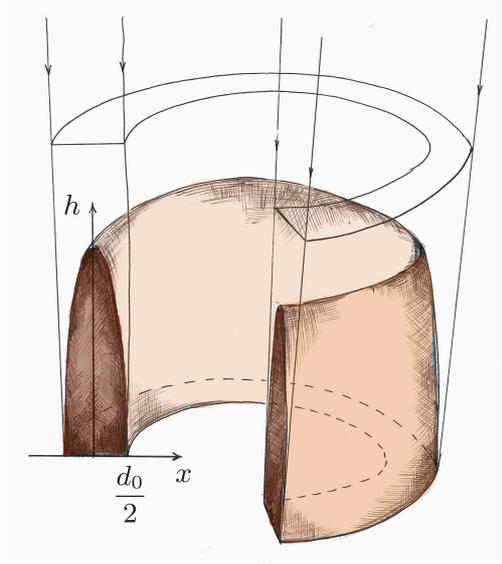} 
\caption{The structure of the accretion column. At the NS surface the cross section of the accretion channel is a thin annular arc of length $l_0$ and thickness $d_0$. The position of the radiation-dominated shock varies inside the accretion 
channel, because the radiation energy density drops sharply towards the column edges. 
As a result the accretion column (i.e. the settling flow) 
height reaches its maximum inside the channel and decreases towards its borders.}
\label{pic:scheme}
\end{figure}

Let us consider an X-ray pulsar accreting at a sufficiently high rate. 
If luminosity  exceeds the critical value $L^*$ \citep{BS1976,2015MNRAS.447.1847M}, the radiation-dominated shock 
is formed in the accreting gas. 
After the shock, the hot gas settles to the NS surface in the magnetically confined accretion column, 
whose height depends on the accretion rate and the magnetic field strength (see Fig.~\ref{pic:scheme}). 
At high accretion rates that are of interest here, 
the optical depth of the settling flow along and across the  magnetic field direction is significantly larger than unity. 
 
Our model is based on a description of optically thick accretion column proposed by \citet{LS1988}. The cross-section of the accretion column is a thin annular arc of width $d_0$ which is much less than its length $l_0$. Then the area of the accretion column base is $S_{\rm D} = l_0d_0$.

The gravitational force, which acts on the gas in the accretion column, will be offset by the radiation pressure gradient only.\footnote{Photon diffusion time from the centre of the column is $t_{\rm diff}=d_0\tau/(2c)\approx 10^{-5}\,{\rm s}$. The matter settling time in the accretion column is $t_{\rm dyn}\gtrsim 7H/v_{\rm ff}(H)\sim 7\times 10^{-4}\,{\rm s}$ (for the accretion column of height $H=R$). As a result, $t_{\rm diff}\ll t_{\rm dyn}$ and one can neglect in the first approximation the effects of plasma movement.} The gas pressure becomes important at virial temperature $\sim 10^{12}$~K at the NS surface, which cannot be reached due to effective electron-positron pair creation and further neutrino cooling at much lower temperatures $\sim 10^{10}$~K \citep{BPS1967,BS1976}. The radiation pressure gradient at the centre of the accretion channel can be presented as follows:
\be \label{dprad}
\frac{dP_{\rm rad}(x=0,h)}{dh} = -\rho\frac{GM}{(R+h)^2},      
\ee
where $M$ and $R$ are NS mass and radius, $h$ is the height above the stellar surface, $\rho$ is the local mass density in the column. 
This equation gives an upper limit of $P_{\rm rad}$, since we ignore the contribution of the gas pressure and deviations 
of the hydrostatic equilibrium due to a subsonic  settlement of the gas in the column. 
This approximation is equivalent to a suggestion that the radial radiation flux at every height of the accretion column 
equals the local Eddington flux 
\be
F_{\rm Edd} = \frac{c}{\kappa_{||}}\, \frac{GM}{(R+h)^2},
\ee
where $\kappa_{||}$ is the opacity along the magnetic field lines. 
Assuming  constant density and taking a zero boundary condition ($P_{\rm rad}(x=0,H)=0$) for the hydrostatic equilibrium equation,
we can get a simple analytical expression for the radiation pressure distribution along the column height
\be\label{eq:prad}
P_{\rm rad} (x=0,h) \approx \rho \frac{GM}{R+H}\,\frac{H-h}{R+h}, 
\ee    
where $H$ is the height of the shock.
Assuming further the optically thick column, we can estimate the flux emergent from the column side at a given height $h$ 
using a plane-parallel diffusion approximation for the radiation transfer across the accretion channel: 
\be\label{rt_prp}
   \frac{dP_{\rm rad}(x,h)}{\d x} = -\rho\,\kappa_{\perp}\,\frac{F_{\perp}(x,h)}{c} \approx -\rho\,\kappa_{\perp}\,\frac{F_{\perp,\rm esc}(h)}{c}\,\frac{2x}{d_0},
\ee
where $x$ is a distance from the centre of the accretion channel (see Fig.\ref{pic:scheme}),  $\kappa_{\perp}$ is the opacity across the $B$-field lines, and $F_{\perp,\rm esc}(h)=F_{\perp}(d_0/2,h)$ is the escaping flux.  
The second equality here was obtained  from the condition $F_{\perp}(0,h) = 0$
and assuming that the flux increases linearly towards the edges of the column (where $x=d_0/2$). 
Then the solution of equation (\ref{rt_prp}) is
\be \label{pradt}
 P_{\rm rad}(x,h) \approx \frac{\,\tau_{0}F_{\perp,\rm esc}(h)}{4c}\,\left(1-\frac{4x^2}{d_0^2}\right),
\ee
where $\tau_0=\rho\,\kappa_{\perp} d_0$ is the optical depth across the accretion channel. 
This solution allows us to find an emergent flux: 
\be \label{f_prp}
F_{\perp,\rm esc}(h) \approx \frac{4c\,P_{\rm rad}(x=0,h) }{\tau_{0}} \approx \frac{4c}{\kappa_{\perp}\,d_0}\, \frac{GM}{R+H}\,\frac{H-h}{R+h},      
\ee
which is much higher than the Eddington flux for the case of tall ($H\gg d_0$) accretion columns
\be
F_{\perp,\rm esc}(h) = 4\,\frac{\kappa_{||}}{\kappa_{\perp}}\,\frac{H-h}{d_0}\,\frac{R+h}{R+H}\,F_{\rm Edd}. 
\ee
The total accretion luminosity of a NS can be estimated by integrating the emergent flux of the surface 
of two columns:
\beq \label{eq:lc}
     L= 4l_0 \int_0^H F_{\perp,\rm esc}(h) \,dh \approx 16\,\frac{l_0}{d_0}\,f\left(\frac{H}{R}\right)\,\frac{c}{\kappa_{\perp}}\, GM  \\ \nonumber
     \approx 38 \left(\frac{l_0/d_0}{50}\right)\,
     \left(\frac{\kappa_{\rm T}}{\kappa_{\perp}}\right)\,f\left(\frac{H}{R}\right)\,L_{\rm Edd}, 
\eeq
where $\kappa_{\rm T} \approx 0.34$\,cm$^2$\,g$^{-1}$ is the Thomson electron scattering opacity of the solar mix plasma and 
\be \label{eq:frh}
f\left(\frac{H}{R}\right)=\log\left(1+\frac{H}{R}\right) -\frac{H}{R+H}.
\ee 
The function $f(x)$ grows quadratically $f(H/R) \sim (H/R)^2$ at small column heights $H \ll R$ and grows logarithmically at $H>R$. 
For the simple analytical estimate (\ref{eq:lc}) we assumed that the optical thickness across the column is constant. 
The dependence of the optical thickness on the height will be taken into account accurately in our numerical model (see Section \ref{sec:ColGeomAndStructure}). 
The maximum X-ray pulsar luminosity can be estimated from equation (\ref{eq:lc}) substituting $H=R$ \citep{BS1976} 
\be    \label{ltwst}  
 L^{**} \!\! = L (H=R) \approx 1.8\times 10^{39}  \!\!  \left( \!\frac{l_0/d_0}{50} \!\! \right)
     \left(\frac{\kappa_{\rm T}}{\kappa_{\perp}}\right) \frac{M}{\msun}{\rm erg\,s^{-1}} .
\ee

We have suggested earlier that the accretion column thickness does not depend on the height above the surface. In fact, the radiation pressure drops down towards the accretion channel edges (see equation (\ref{pradt})). As a result, the radiation pressure cannot support the same column height all over the accretion channel. The height has its maximum in the middle of the channel and decreases towards its edges. Therefore, the column thickness has to decrease with the height as well (see Fig. \ref{pic:scheme}). 
Taking the radiation pressure $P_{\rm rad}(x,h=0)$ at the bottom of the accretion column 
we can evaluate the column height $H_x$ (i.e. the position of the radiative shock) as for a given $x$  
using equation (\ref{eq:prad})  where $H$ is replaced with $H_x$. This height drops 
from the maximum value $H$ at the centre to the zero value at the column edge:  
\be \label{hc}
\frac{H_x}{R+H_x} = \frac{H}{R+H}\,\left(1-\frac{4x^2}{d_0^2}\right). 
\ee  
Therefore, the radial cross section of homogeneous column of a small height ($H \ll R$) 
has a roughly parabolic form (Fig.~\ref{pic:scheme}).

\subsubsection{Consistency check} 

Let us estimate the values of the basic parameters in order to justify the proposed model. Let us consider a NS of mass $M=1.4\msun$ and radius $R=10\,{\rm km}$ accreting at a rate $\dot M \approx 5\times 10^{18}$~g\,s$^{-1}$ 
(which gives accretion luminosity $L=10^{39}\,$\ergs). 
We take the column base area $S_{\rm D} = 5\times 10^{9}$~cm$^2$ and its geometrical size are $l_0=5\times 10^5$~cm 
and $d_0=10^4$~cm, 
which are expected for a given mass accretion rate for standard magnetic field of $B=10^{12}$~G \citep{BS1976}. 
We also assume here $\kappa_{\perp} = \kappa_{\rm T}$. 

According to equation (\ref{eq:lc}) the corresponding height of the accretion column is $H\approx 0.75\,R$. The free-fall velocity at the height $H$ above the NS surface $v_{\rm ff} =\sqrt{2GM/(R+H)}\approx 1.5\times 10^{10}$~cm\,s$^{-1}$ and the  mass density for given mass accretion rate $\rho_{\rm ff}\approx 3.5\times 10^{-2}$~g\,cm$^{-3}$. 
Because the matter is decelerated by the radiation-dominated shock at the top of the accretion column \citep{BS1976}, 
the plasma velocity falls down to $v = v_{\rm ff}/7$ \citep[see, e.g.][]{1982SvAL....8..330L,1998ApJ...498..790B}, 
and the mass density goes up to $\rho \approx 0.245$~g\,cm$^{-3}$. 
As a result the Thomson optical thickness across the accretion channel is $\tau_0 = \rho \kappa_{\rm T} d_0 \approx 500$, 
which means that the accretion column is optically thick as it was assumed earlier.

The radiation pressure  and the emergent side flux given by equations (\ref{eq:prad}) and (\ref{f_prp}),
respectively, near the NS surface are $P_{\rm rad}\approx 3.5\times 10^{19}$~dyn\,cm$^{-2}$ and $F_{\perp} \approx 8\times 10^{28}\,$\ergs\,cm$^{-2}$. 
Because the column is optically thick we can assume thermodynamical equilibrium deep inside it 
and evaluate the plasma temperature from the radiation field energy density:
\be \label{td}
 P_{\rm rad} \approx \frac{\varepsilon_{\rm rad}}{3} \approx \frac{aT^4}{3},
\ee
where $a$ is the radiation constant. This relation gives $T \approx 3.4\times 10^8$~K or $kT \approx 30$~keV. Therefore, the gas pressure $P_{\rm g} = \rho kT/(\mu_{\rm s} m_{\rm H}) \approx 5\times 10^{15}$~dyn\,cm$^{-2}$ is much smaller than the radiation pressure $P_{\rm rad}$, as we assumed earlier (here  $\mu_{\rm s} = 0.62$ is the mean particle weight for fully ionized solar mix plasma). 
The effective temperature corresponding to the escaping flux is determined by the relation 
$F_{\perp} = \sigma_{\rm SB}T_{\perp}^4$  is $T_{\perp} \approx 1.9\times 10^8$~K or $kT_{\perp} \approx 16$~keV. 
This value is close to the cut-off energy observed in the spectra of bright X-ray pulsars, 
especially if the gravitational redshift is taken into account \citep{2005AstL...31..729F}. 
We see thus that the assumption made constructing the accretion column model are reasonable and 
the model is  self-consistent. We consider this analytical model as a base for further numerical studies.

In the analytical estimates, we have considered a homogeneous column with a constant cross-section area. 
For the numerical model described in the next section, 
we account for the column inhomogeneity and take the exact opacity of highly-magnetized plasma.

\subsection{Geometry and structure of the accretion column}
\label{sec:ColGeomAndStructure}

Using the approach described  in Section~\ref{sec:AnalyticalEstimates}, we can construct a numerical model of the accretion column. Global input model parameters are the NS mass $M$ and radius $R$, the mass accretion rate $\dot M$, 
and the polar magnetic field strength $B$. The magnetic field is assumed to be dipole with the magnetic
field strength decreasing with distance as
\be \label{bh}
  B_h \approx B\,\left(\frac{R+h}{R}\right)^{-3} = B\, S_h^{-1},
\ee  
where $S_h$ is the cross-section area of the accretion channel at a given height $h$ above the stellar surface. 
The width and the length of the accretion channel depends on  height as follows:
\be
d \simeq d_0\left(\frac{R+h}{R}\right)^{3/2},\quad l\simeq l_0\left(\frac{R+h}{R}\right)^{3/2}.
\ee
The geometrical parameters of accretion column base $l_0$ and $d_0$ can be derived from the global parameters of the model (see Sect.~\ref{sec:AccretionColumnBases}).
It has to be mentioned that the thickness of a sinking region of the accretion column $d_{h}$ at some non-zero height $h$ is always smaller than its base thickness $d_0$ because of the parabolic shape of the shock surface. 
The column thus is surrounded by almost free-fall accreting matter of geometrical thickness $\sim(d-d_{h})/2$, 
which affects the distribution of the emergent radiation over the directions \citep{LS1988,Poutanen2013}.

We consider a two-dimensional radial cross section of the column using two geometrical coordinates $h$ and $x$ (Fig.~\ref{pic:scheme}). The column structure is determined by a set of equations including local mass conservation law and hydrostatic equilibrium. The mass conservation law can be expressed as follows:   
\be \label{m_cnsv}
  \frac{\dot M}{2\,S_{\rm D}} = \rho v ,
\ee
where $\rho$ and $v$ are local mass density and velocity correspondingly. 
The exact local values of $\rho$ and $v$ cannot be calculated without solving the full set of radiation-hydrodynamic equations. The velocity profiles in a sinking region were calculated for particular cases \citep{1981A&A....93..255W} and can be approximated by power law $v\approx h^\xi$ with the exponent $\xi\sim 1$ for relatively low accretion luminosity $\sim 10^{37}\,{\rm erg\,s^{-1}}$ and with higher $\xi\in[1;5]$ depending on the accretion channel thickness, opacity in a sinking region and magnetic field structure. For the case of high accretion column ($H\sim 3.5R$) and $\sigma\equiv\sigma_{\rm T}$ the velocity profile was calculated to be $v\approx h^5$ \citep{BS1976}.  Thus, we assume that the velocity profile in the accretion column below the shock zone is given by a power law and discuss influence of the parameter $\xi$ on the solutions (see Section\,\ref{sec:Results}).
The velocity at the NS surface is equal to zero, while the plasma velocity at the top of the column just below the shock is $v(H)=v_{\rm ff}(H)/7$ \citep[see][]{1982SvAL....8..330L,1998ApJ...498..790B}, where $v_{\rm ff}(H)=\sqrt{2GM/(R+H)}$ is a free-fall velocity at height $H$ above the NS surface. 

According to equation (\ref{dprad}), the local radiation pressure $P_{\rm rad}$ can be presented as follows:
\be \label{pradf}
P_{\rm rad}(x,h)\approx \int\limits_{h}^{H_x}
\rho\,\frac{GM}{(R+y)^{2}}\,\d y 
+ \frac{2}{3}\,\frac{F_{\rm Edd}(H_x)}{c},
\ee
where we used standard boundary condition for the radiation density $\varepsilon_{\rm rad} = 2F/c$ and the approximation of thermodynamic equilibrium (\ref{td}).
The connection between radiation pressure $P_{\rm rad}$ at the centre of the accretion channel and the emergent radiation flux $F_{\perp,\rm esc}$ 
is determined approximately by the radiative transfer equation (\ref{rt_prp}). However, for the accurate solution one has to account that the opacity $\kappa_{\perp}$ depends on the local column conditions such as the temperature and the $B$-field strength 
(see details in Section \ref{sec:Cross-Section}). Equation\,(\ref{rt_prp}) can be transformed to 
\be \label{rt_prpf}
P_{\rm rad}(x,h) \approx \frac{2\rho}{d_h}\,\frac{F_{\perp,\rm esc}(h)}{c}\,
\int\limits_{x}^{d_h/2}\,\kappa_{\perp}\,z\,\d z + \frac{2}{3}\,\frac{F_{\perp,\rm esc}(h)}{c},
\ee
where the similar surface boundary condition like in equation (\ref{pradf}) was used. 
Because $\kappa_{\perp}$ depends on the local temperature which is evaluated from $P_{\rm rad}(x,h)$ (see eq.\,(\ref{td})), 
equation (\ref{rt_prpf}) could be solved iteratively, if one knows the radiation pressure $P_{\rm rad}$ at the centre of the accretion channel.  
Then the flux emergent from the wall of the column $F_{\perp,\rm esc}$ can be found for the known column thickness $d_h$ 
and plasma mass density $\rho$ at every height. The thickness and the mass density are found iteratively as well (see the iterative scheme below). 

The described model could be incorrect for the case of extremely heigh accretion columns ($H \gtrsim R$) due to geometrical reason, namely due to curvature of dipole magnetic field lines, when the magnetic field and gravity are not aligned and condition given by eq.\,(\ref{dprad}) is not valid. It means that we cannot find any solution at some high luminosities when $H>R$. 
Fortunately, the approximate formulae (\ref{eq:lc}) and (\ref{eq:frh}) show that the luminosity saturates at $H \approx R$ and we still can obtain a 
sufficiently good estimation of the maximum X-ray pulsar luminosity for given NS mass and radius and the magnetic field strength.

\subsection{Scattering cross-section}
\label{sec:Cross-Section}

In order to compute the accretion column luminosity or the accretion column height for a given luminosity, 
we need to find the mean opacity across $\kappa_{\perp}$ and along $\kappa_{||}$ the magnetic field. 
We assume that the plasma in the column is fully ionized  and ignore bremsstrahlung opacity. 
Thus, the plasma opacity is determined by Compton scattering only. 
Compton scattering cross-section in case of high magnetic field depends strongly on the photon energy, polarization state and 
photon momentum direction \citep{DH1986,Mushtukov2015PR}. 
In the case of optically thick plasma, in diffusion approximation, 
we only require the Rosseland mean opacities in the two directions: 
\be
     \kappa_{\perp,||} = \overline\sigma_{\perp,||}\,\frac{n_{\rm e}}{\rho} = \frac{\overline\sigma_{\perp,||}}{\mu_{\rm e}m_{\rm H}},
\ee 
where $\overline\sigma_{\perp,||}$ are the Rosseland mean Compton scattering cross-sections across and along magnetic field (see below),
$\mu_{\rm e} = 1.17$ is the mean number of nucleons per electron for fully ionized solar mix plasma.
For the angular-dependent case, the cross-section along  and across the $B$-field are
\citep{1992AcA....42..145P,Putten2013,2015MNRAS.447.1847M}
\be\label{eq:ross_ang_par} 
\frac{1}{\overline{\sigma}_{\parallel,j}}=\frac{\displaystyle\int\limits_{0}^{\infty} 
  \frac{\d B_{\rm E}(T)} {\d T}\d E  \int\limits_{0}^{1}\d\mu\,3\mu^2 
  \frac{1}{\sigma_{j}(E,\mu ,T)}}{\displaystyle\int_{0}^{\infty}\frac{\d B_{\rm E}(T)} {\d T}\d E}, 
\ee 
and  
\be 
\frac{1}{\overline{\sigma}_{\perp, j}}=
\frac{\displaystyle\int\limits_{0}^{\infty}\frac{\d B_{E}(T)}{\d T}\d E
\int\limits_{0}^{\pi}\d \varphi\int\limits_{0}^{1}\d\mu_{\rm n}\frac{3}{\pi}\mu^2_{\rm n}\frac{1}{\sigma_j(E,\mu,T)}}
{\displaystyle\int_{0}^{\infty}\frac{\d B_{E}}{\d T}\d E},
\ee
respectively. 
Here $T$ is the local electron temperature, $B_{E}(T)$ is the Planck function, $E$ is a photon energy, 
$\mu$ is a cosine of the angle between photon momentum and the $B$-field direction and 
$\mu_{\rm n}$ is a cosine between the momentum and perpendicular to the field direction ($\mu=\sqrt{1-\mu_{\rm n}^2}\cos\varphi$). 
Index $j$ correspond to a given photon polarization state ($j=1$ for $X$- and $j=2$ for $O$-mode). 
In case of mixed polarization the effective cross-section can be estimated as follows
\be
1/\overline{\sigma}=f/\overline{\sigma}_1+(1-f)/\overline{\sigma}_2,
\ee
where $f$ is a fraction of radiation in the $X$-mode \citep{2015MNRAS.447.1847M}. We assume $f=1$ for further calculations. 
It is a rather good approximation because most of  the radiation is expected to be in $X$-mode 
due to the effective swapping of photons from $O$- to $X$-mode \citep{1995ApJ...448L..29M,Mushtukov2012}.
In case of sufficiently strong magnetic field the typical photon energy $E$ can be much lower than the cyclotron one $E_{\rm cycl}\simeq 11.6\,B_{12}\,{\rm keV}$. In case of $E<E_{\rm cycl}$ the scattering cross-sections can be approximated as $\sigma_{\perp}(E)\simeq \sigma_{\rm T}(E/E_{\rm cycl})^2$ and $\sigma_{\parallel}(E)\simeq \sigma_{\rm T}[(1-\mu^2)+\mu^2(E/E_{\rm cycl})^2]$ for $X$- and $O$-mode respectively \citep{1971PhRvD...3.2303C}. As a result the effective cross-section for extremely strong $B$-field tends to be significantly lower that the Thomson value.

The computed values of $\sigma_{(\perp,||)(1,2)}$ were tabulated for 40 values of the magnetic field strength  distributed logarithmically  
in the range  $10^{11}-10^{15}$~G. The opacities were calculated for 300 values of the temperature in the range $10^6 - 10^9$~K for every magnetic field strength $B$. 
The variable steps were specially adopted for more detailed description in the temperature region near corresponding electron cyclotron energy.

\subsection{Accretion column base geometry}
\label{sec:AccretionColumnBases}

It is assumed that the accretion flow is interrupted at the magnetospheric radius due to the sufficiently strong magnetic field pressure. The inner radius of the accretion disc depends on the magnetic field strength and its structure, mass accretion rate and the feeding mechanism of the pulsar. 
It can be estimated with the following expression \citep{GL1978,GL1979,GL1992}
\be
R_{\rm m}=7\times 10^{7}\Lambda\, m^{1/7}R^{10/7}_{6}B^{4/7}_{12}L^{-2/7}_{39}\,\, \mbox{cm}, 
\ee
where $\Lambda$ is a constant depending on the accretion flow geometry ($\Lambda=1$ for the case of spherical accretion and $\Lambda<1$ for the case of accretion through the disc, with $\Lambda=0.5$ being a commonly used value), $m=M/\msun$ 
and we use notations in cgs units for other variables $Q=Q_x 10^x$.

Accretion channel geometry is defined by the interaction between the accretion disc and NS magnetosphere. 
The thickness of the accretion channel depends on the penetration depth $\delta$ of the accretion disc into the NS magnetosphere. 
It is expected to be of the order of the disc scale-height at the inner edge  $H_{\rm d}$. 
The accretion disc structure at the inner edge depends on the $B$-field strength and the luminosity. 
We evaluate $H_{\rm d}$ using  the \citet{SS1973} model with a slightly different vertical averaging 
and with the accurate Kramer opacity \citep*{Sul2007}. 
The disc can be divided into three zones according to the dominating pressure and opacity sources. 
The gas pressure and the Kramer opacity dominate in the outer regions (C-zone), the gas pressure and the electron scattering dominate in the intermediate zone (B-zone) and  the radiation pressure may dominate in the inner zone (A-zone). 
The boundaries between A- and B- and between B- and C- zones are expected to be at 
$r_{\rm AB}\approx 2.2\times 10^8 L_{39}^{16/21}m^{-3/7}R_{6}^{16/21}$~cm and 
$r_{\rm BC}\approx 9\times 10^8 L_{39}^{2/3}m^{-1/3}R_{6}^{2/3}$~cm, respectively. 
The disc structure at the inner edge is  defined then by the magnetospheric radius $R_{\rm m}$. 

A relative disc scale-height at radius $r$ for the C-zone, where accretion disc is expected to be interrupted in case of 
relatively low mass accretion rate, is
\beq\label{SS_H}
\left(\frac{H_{\rm d}}{r}\right)_{\rm C}\approx 0.056 \,\alpha^{-1/10}\, L_{39}^{3/20}\, m^{-21/40}\,R_6^{3/20}\,r_8^{1/8}. 
\eeq
Assuming that matter leaves the disc at magnetosphere radius and follows the dipole magnetic field lines down to the polar caps of the NS we get the ratio $l_0/d_0$ and the area $S_{\rm D}$. For the case of a truncation in the C-zone:
\be \label{ldc}
\left(\frac{l_0}{d_0}\right)_{\rm C}\approx 2\pi\frac{R_{\rm m}}{H_{\rm d}}\simeq 95\,\Lambda^{-1/8}m^{71/140}B_{12}^{-1/14}L_{39}^{-4/35},
\ee
and corresponding area of the accretion channel base 
\be
S^{(\rm C)}_{\rm D}\approx 6.6\times 10^{9}\,\Lambda^{-7/8}L^{2/5}_{39}B^{-1/2}_{12}m^{-13/20}R^{19/10}_{6}\,\mbox{cm}^2. 
\ee
A relative disc scale-height in a relatively narrow B-zone is
\beq\label{SS_H}
\left(\frac{H_{\rm d}}{r}\right)_{\rm B}\approx 0.064 \,\alpha^{-1/10}\,L_{39}^{1/5}\, m^{-11/20}\,R_6^{1/5}\,r_8^{1/20},
\eeq
and the ratio $l_0/d_0$ can be estimated as follows
\be \label{ldb}
\left(\frac{l_0}{d_0}\right)_{\rm B}\approx 80\,\Lambda^{-1/20}m^{19/35}B_{12}^{-1/35}L_{39}^{-13/70}.
\ee
The accretion channel base area in this case is
\be
S^{(\rm B)}_{\rm D}\simeq 7\times 10^{9}\,\Lambda^{-19/20}L^{19/70}_{39}B^{-41/70}_{12}m^{-24/35}R^{129/70}_{6}\,\mbox{cm}^2. 
\ee

In the case of sufficiently high luminosity, $L_{39}\gtrsim 0.34\,\Lambda^{21/22}B_{12}^{6/11}m^{6/11}$, 
the structure of the accretion disc zone A is defined mostly by the radiation pressure and 
the disc scale-height is independent of radius: 
\citep{SS1973,Sul2007}:
\be
H_{\rm d,A}\approx\kappa_{\rm T}\frac{3}{8\pi}\frac{\dot{M}}{c}\simeq 10^{7}\,\frac{L_{39}R_{6}}{m}\,\mbox{cm} 
\ee
and therefore 
\be \label{lda}
\left(\frac{l_0}{d_0}\right)_{\rm A}\approx 2\pi\frac{R_{\rm m}}{H_{\rm d}}\simeq 44\,\Lambda\,m^{8/7}R^{3/7}_{6}B^{4/7}_{12}L^{-9/7}_{39}.
\ee
The corresponding area of the accretion channel base
\be
S^{(\rm A)}_{\rm D}=1.3\times 10^{10}\,\Lambda^{-2}L^{11/7}_{39}B^{-8/7}_{12}m^{-9/7}R^{8/7}_{6}\,\mbox{cm}^2. 
\ee

If the magnetic dipole is inclined with respect to the orbital plane, the expression for the hot spot area and the ratio $(l_0/d_0)$ would be different and the spot would have a shape of an open ring. We use an additional parameter $l^{*}_{0}/l_0$, which shows what part of the full ring length $l_0$ is filled 
with accreting gas. 
In numerical calculations we take $l^{*}_{0}/l_0\geq 0.5$. 
It is also natural to assume that the ratio $l^{*}_{0}/l_0$ is higher for higher mass accretion rates.

\subsection{Computation scheme}
\label{sec:Iterr}

A general computation scheme consists of a few steps. The main task is to find the central accretion column height $H$, which corresponds to a given X-ray pulsar luminosity $L$ (i.e. to a given mass accretion rate $\dot M$). The parameters of the model are NS mass $M$ and radius $R$, surface magnetic field strength $B$, parameter $\xi$, which determines the velocity profile in the sinking region, the polarization composition of radiation given by parameter $f$ (see Section\,\ref{sec:Cross-Section}) and the ratio $l^*_0/l_0$, which gives what part of the full ring is filled with the accreting matter (see Section\,\ref{sec:AccretionColumnBases}). The accretion column base geometry ($l_0$, $d_0$) is defined by $L$ and $B$ (see Section\,\ref{sec:AccretionColumnBases}). The accretion column height and the structure are calculated by double iteration scheme. The external iteration finds the accretion column height for a given set of parameters by the dichotomy method, while the internal iteration calculates the luminosity for current accretion column height and given base geometry.
We make a first guess of the accretion column height using approximation (\ref{eq:lc}) and assuming $\kappa_{\perp} = \kappa_{\rm T}$. The steps of the iterative scheme are: 
\begin{enumerate}
\item The temperature inside the sinking region $T(x,h)$ is calculated analytically using equations (\ref{eq:prad}), (\ref{pradt}), (\ref{f_prp}) and (\ref{td}) as a first guess and recalculated iteratively further (see below).
\item For the current temperature distribution we compute the opacity $\kappa_{\parallel,\perp}(x,h)$ (see Section\,\ref{sec:Cross-Section}). 
\item For the current accretion column height $H$ we compute the central radiation pressure ($x=0$) using formula (\ref{pradf}) as a function of height $h$ inside the sinking region, i.e. we get $P_{\rm rad}(0,h)$ for $h\in [0;H]$. Formally, formulae (\ref{m_cnsv})--(\ref{pradf}) lead to the infinite values of the mass density and the radiation pressure at $h \rightarrow 0$. 
It means that formula (\ref{pradf}) is inaccurate at the very bottom of the accretion column. The total pressure becomes 
higher than the magnetic pressure $(P_{\rm rad} +P_{\rm g}) > B^2/4\pi$ and plasma spreads over the NS surface. The height $h_0$, where it happens, is taken as the base of the accretion column.\footnote{The height $h_0$ depends on the mass accretion rate, $B$-field strength and velocity profile in the sinking region and takes on a value $\sim (0.1 - 100)\,{\rm cm}$, which is $\ll d_0$ and does not affect the luminosity calculated for given accretion column height.}
\item Then we find a dependence of the thickness of a sinking region $d_h$ on height $h$. 
For that, we first compute $F_{\perp,\rm esc}(h=h_0)$ using formula (\ref{rt_prpf}) from the already known $P_{\rm rad}(0,h_0)$, current opacity $\kappa_\perp(x,h)$ and known thickness at the base $d_{h_0}=d_0$. 
Then using the same formula (\ref{rt_prpf}) and already known $F_{\perp,\rm esc}(h=h_0)$ we compute the radiation pressure at the base $h=h_0$ from the middle of the column to the edge, i.e. we get $P_{\rm rad}(x,h=h_0)$ for $x\in[0;d_0/2]$. The evaluated radiation pressure at the accretion column base $P_{\rm rad}(x,h=h_0)$ is used to get the column height $H_x$ at distance $x$ from the center of the channel (see equation (\ref{pradf})). Then we get $P_{\rm rad}(x,h)$ using again equation (\ref{pradf}). Reversing the function $H_x$ we finally get the dependence of a sinking region width $d_h$ on the height $h$.
\item As soon as the central radiation pressure $P_{\rm rad}(0,h)$ and the thickness of the sinking region $d_h$ are known at any height, 
we use equation (\ref{rt_prpf}) and get the emergent flux $F_{\perp,\rm esc}(h)$ as a function of $h$.
\item We get the resulting X-ray pulsar luminosity for a given accretion column height $H$ and the temperature structure
\be 
L' \simeq 4\,l_0\,\int\limits_{0}^{H} \left(\frac{R+h}{R}\right)^{3/2}F_{\perp,{\rm esc}}(h)\d h.
\ee
\item Using computed $P_{\rm rad}(x,h)$ we get new temperature distribution in the accretion column $T(x,h)=(3\,P_{\rm rad}(x,h)/a)^{1/4}$. Then we return to step (ii), recalculate the luminosity $L'$ for a given column height $H$ and new temperature $T(x,h)$ and continue this procedure until we get the accretion luminosity $L'(H)$ with the relative accuracy of 1 per cent.
\item If the luminosity $L'(H)$ does not coincide with the given accretion luminosity $L$ we get the next guess of the accretion column height and return to step (i). We continue the iteration procedure until the relative accuracy of 1 per cent is reached. 
\end{enumerate}
Finally, we get the height and the structure of the accretion column for a given mass accretion rate.

\section{Numerical results}
\label{sec:NumericalResults}

\subsection{Accretion column properties}
\label{sec:NumRes}

There are four parameters (in addition to the NS mass $M$ and radius $R$) defining the accretion column height and structure. 
These are the surface polar NS magnetic field strength $B$, the mass accretion rate $\dot{M}$ or the accretion luminosity $L=GM\dot{M}/R$, 
and the geometrical sizes of the accretion column base, $d_0$ and $l_0$. 
Here we investigate how these parameters affect the properties of the accretion column. 
In order to separate the effects caused by different reasons, we will assume a fixed accretion column base geometry with the following parameters $l_0=7.7\times 10^5$~cm $d_0=1.3\times 10^4$~cm in this section only. 
The self-consistent results, where the dependence of accretion channel geometry on the mass accretion rate is taken into account, 
will be discussed in Section~\ref{sec:Results}.
 
Two main parameters that  determine the global properties of the accretion column 
are the local mass accretion rate $\dot M/2S_{\rm D}$ and the 
effective scattering cross-section across the accretion channel $\sigma_{\perp}$. 
The vertical temperature structure in the centre of sinking region of the accretion column for a given column height depends 
on the local mass accretion rate only. The temperature structure across the accretion channel and, therefore, the emergent flux at given height is 
determined by the local effective scattering cross-sections $\sigma_{\perp(1,2)}$ (see Section \ref{sec:Cross-Section}) 
and the column density  $\int_0^{d_0} \rho dz$ across the channel. 
The stronger the field, the smaller the scattering cross-section and the smaller the optical thickness across the column. 
As a result, the higher the $B$-field, the higher the emergent flux $F_\perp$ for the fixed temperature in the column centre and the column density. 
Global properties of a self-consistent accretion column model, which have to be constructed in an iterative way, are determined by these facts.

\begin{figure}
\centering
\includegraphics[width=0.9\linewidth]{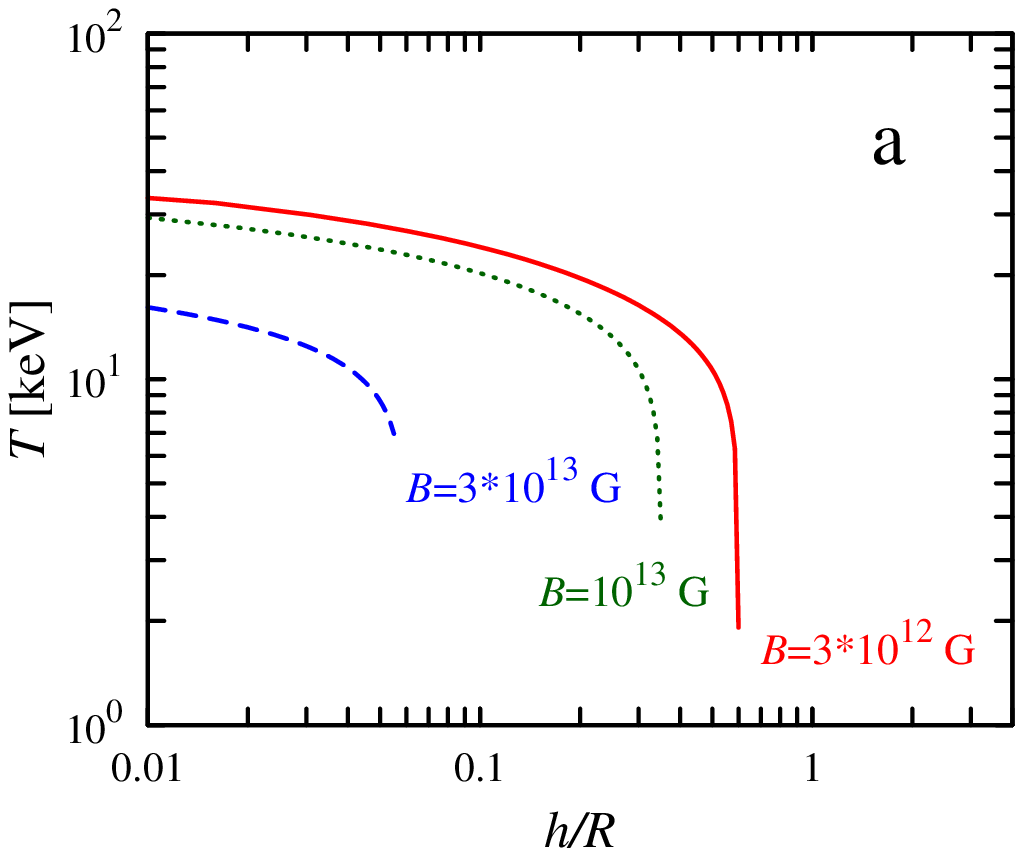}
\includegraphics[width=0.9\linewidth]{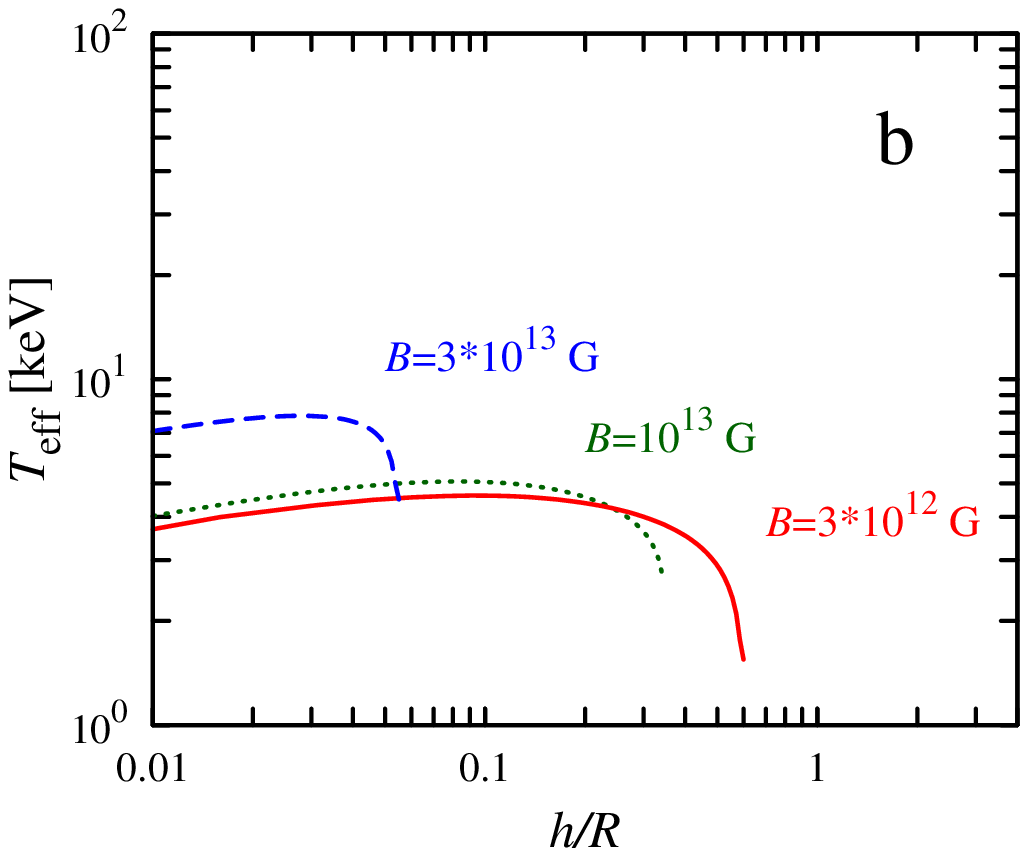}
\includegraphics[width=0.9\linewidth]{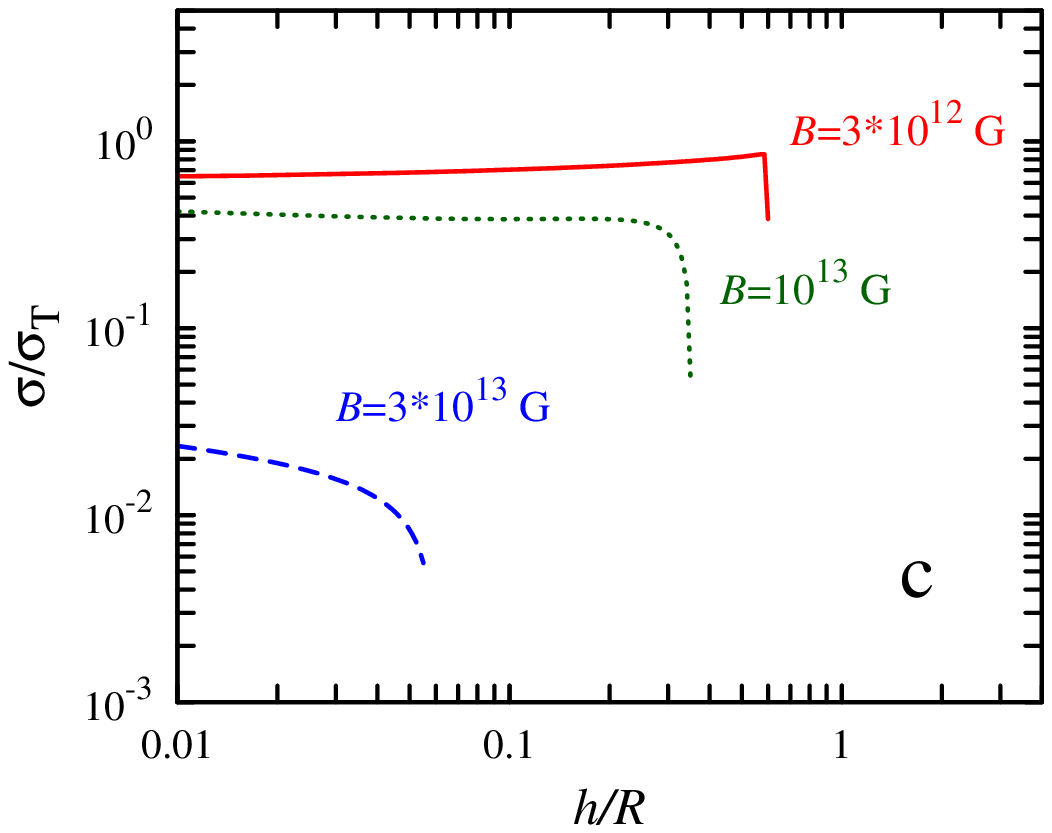}
\caption{Theoretical dependences of (a) the internal temperature, 
(b) the effective surface temperature, and (c) the effective scattering cross-section on the $h/R$ ratio for $L=10^{39}\,$\ergs\
and fixed accretion column base geometry: $l_0=7.7\times 10^5$~cm $d_0=1.3\times 10^4$~cm. 
Red solid, green dotted, and blue dashed lines correspond to $B=3\times 10^{12}$, $10^{13}$ and $3\times 10^{13}$~G,
respectively.
Parameters: $M=1.4 \msun, R=10^{6}$~cm, $\xi=1$, $f=1$.}
\label{pic:theor}
\end{figure}

Let us focus on the dependence of accretion column properties on the magnetic field strength and consider accretion column models corresponding to three different surface $B$-field strength and fixed accretion luminosity $L=10^{39}\,$\ergs\ (see Fig.\,\ref{pic:theor}). The higher field strength leads to lower effective scattering cross-section (Fig. \ref{pic:theor}\,c). Therefore, the emergent flux at a given height ($F_\perp =\sigma_{\rm SB}T_{\rm eff}^{4}$) is higher (Fig. \ref{pic:theor}\,b). Smaller column at high $B$-field provides the same luminosity as higher column at relatively low field strength. On the other hand, the increased leaking of thermal energy makes the central column temperature lower in case of high $B$-field (Fig.\,\ref{pic:theor}\,a) and supporting of the accretion column is less effective. Then the accretion column is also smaller in case of higher field strength for fixed mass accretion rate. We note, that the significant dependence of accretion column properties on the magnetic field strength takes place for sufficiently high magnetic field strengths ($B > 10^{12}$~G) only, when most energy is radiated below the electron cyclotron energy $E_{\rm cyc}=11.6\,B_{12}$~keV 
and the effective electron scattering cross-section drops significantly below $\sigma_{\rm T}$ (Fig.\,\ref{pic:theor}\,c). 
As $B \lesssim 10^{12}$~G, the properties of sufficiently bright accretion columns become almost independent on the magnetic field strength. 

\begin{figure*}
\centering 
\includegraphics[width=8cm]{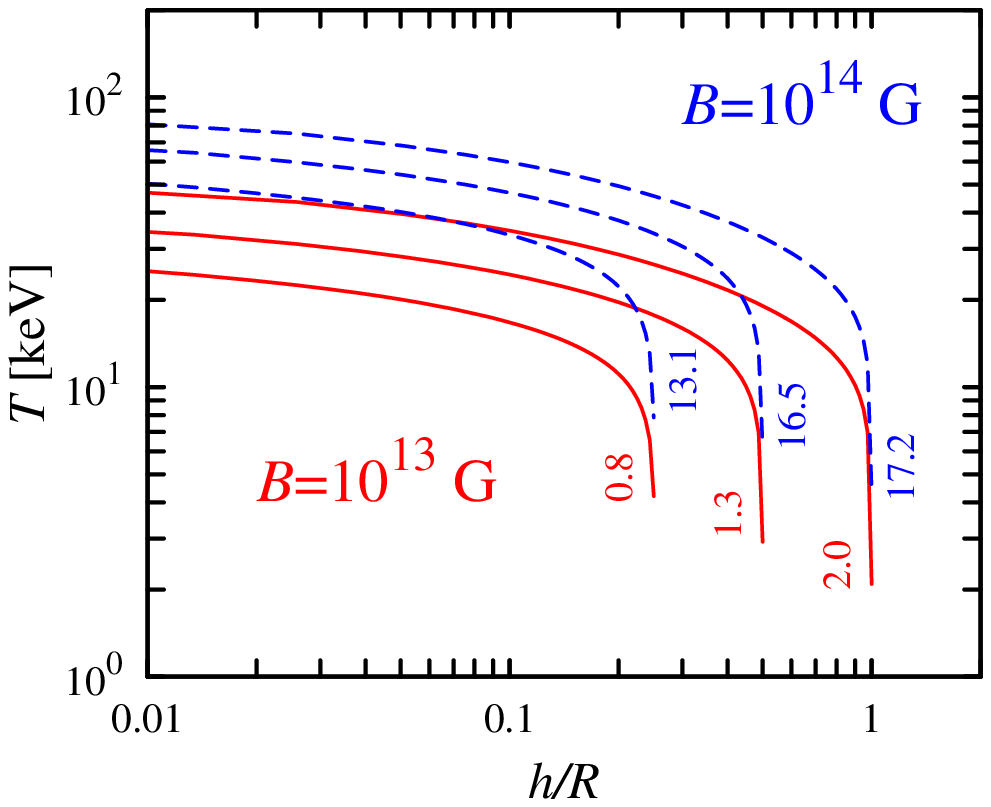} 
\includegraphics[width=8cm]{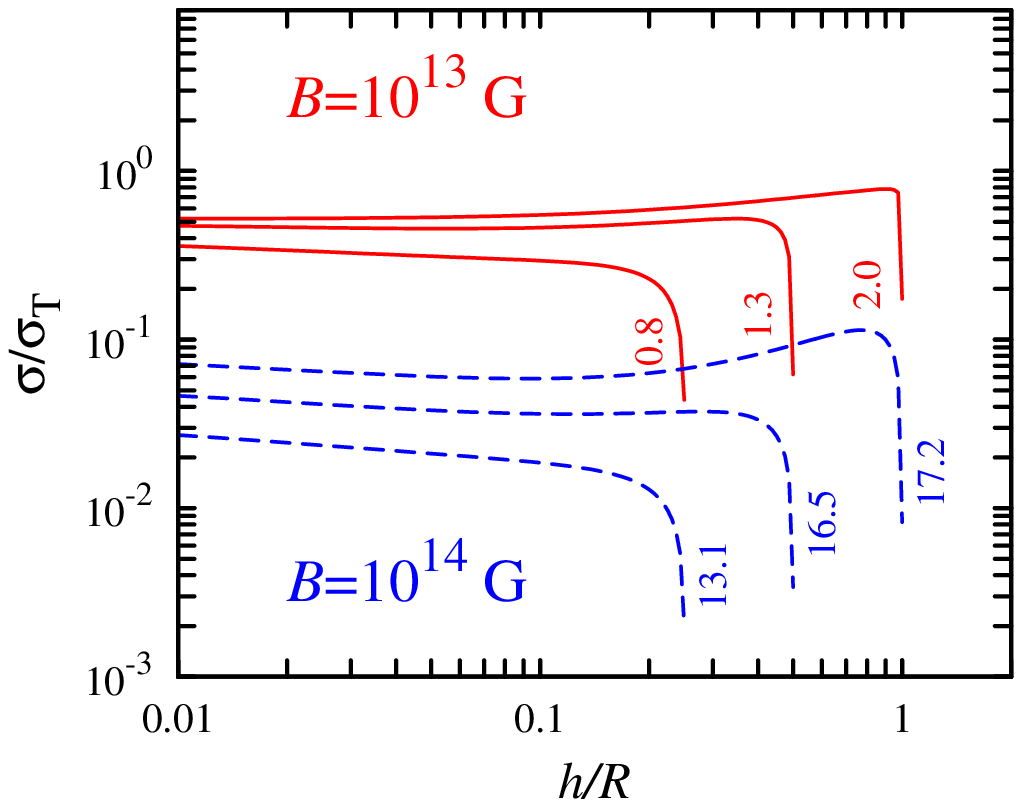} 
\caption{The internal temperature and the effective scattering cross-section as a function of height. Different curves correspond to
various accretion column height: $0.25R,\,0.5R,\,R$ and different luminosities, which depend on the field strength. The accretion column base geometry is fixed: $l_0=7.7\times 10^5$~cm, $d_0=1.3\times 10^{4}$~cm. Red solid lines  correspond to 
the case of surface magnetic field strength $B=10^{13}$~G, while blue dashed lines correspond to $B=10^{14}$~G. 
The corresponding luminosities (in units of $10^{39}\,$\ergs) are marked next to each curve. 
The higher the accretion column, the higher the internal temperature, the higher effective scattering cross-section for a 
given height above the NS surface. Parameters: $M=1.4 \msun, R=10^{6}$~cm, $\xi=1$, $f=1$.}
\label{pic:T_and_sigma} 
\end{figure*}

The effective temperature of the accretion column is not a monotonic function of height and reaches its maximum value above the NS surface (Fig.~ \ref{pic:theor}b). 
On the contrary, the internal temperature decreases monotonically with height (Fig.~\ref{pic:theor}a), according to the basic principles described above. 
The value of internal temperature on the top of the column is defined by the boundary condition in equation (\ref{pradf}). 
The internal central temperature is higher than the effective temperature at a given height and this difference is defined by the optical thickness 
across the accretion column. 
The optical depth across the column decreases with the height because of decreasing of geometrical thickness 
(see Section~\ref{sec:AnalyticalEstimates}) 
and overcompensates the drop of the internal temperature. This causes the increase of the effective temperature with height.     

We note that the difference between the effective and the internal temperatures is larger for a relatively low $B$-field strength at a fixed luminosity, 
because of higher electron scattering opacity. 
These temperatures become close to each other as soon as the optical thickness across the column becomes close to unity. It happens when the accretion luminosity is close to the critical luminosity value $L^*$ for a given magnetic field strength \citep{2015MNRAS.447.1847M}, which defines the minimum luminosity when the accretion column exist. 

It was shown above that the column reaches its maximum luminosity when the column height $H \approx R$. 
Therefore, a stronger magnetic field allows X-ray pulsars to reach higher luminosities, i.e. the accretion column of a fixed height will be more 
luminous when the magnetic field is stronger. We illustrate this statement by a few accretion columns with the same accretion column base geometry 
and different luminosities, for two fixed magnetic field strengths: $10^{13}$ and $10^{14}$~G (see Fig.\,\ref{pic:T_and_sigma}). 
It is clear that the accretion columns of similar heights but different field strength produce luminosities that may differ by an order of magnitude.

The dependences of column properties on the geometrical parameters are much simpler and we do not illustrate it in details. Decrease of the column base thickness $d_0$ and the increase of column base length $l_0$ lead to lower accretion column heights at a  given mass accretion rate and  
the $B$-field strength. In fact, the geometrical parameters are not independent, they depend on the magnetospheric radius $R_{\rm m}$ and the 
accretion disc scale-height at its inner edge. Both of them depend on the mass accretion rate and the magnetic field strength on the NS surface. We consider this problem in detail in the next section.

\begin{figure}
\centering 
\includegraphics[width=8.5cm]{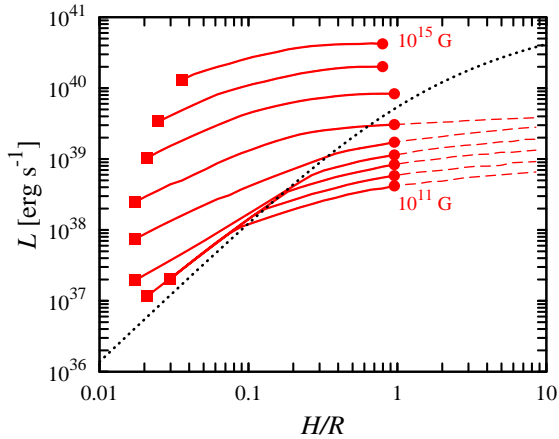} 
\caption{The accretion column luminosity as a function of its height. 
Red solid lines correspond to various surface magnetic field strength: 
$B=10^{11}, 3\times 10^{11}, 10^{12}, 3\times 10^{12}, 10^{13}, 3\times 10^{13}, 10^{14}, 3\times 10^{14}$ and $10^{15}$~G. 
The red circles at the right end of the curves indicate the maximum accretion luminosity values and the 
corresponding accretion column height. 
The red squares indicate the minimum critical luminosity $L^*$ when the accretion column exists. 
The saturation of the luminosity is caused by two different factors: 
a switch of the accretion disc behaviour on its inner radius for $B\lesssim 10^{13}$~G  
and by increasing of the effective scattering cross-section for $B\gtrsim 10^{13}$~G,  
which makes the column growth ineffective at some point. 
The black dotted line shows the approximate dependence (\ref{eq:frh}) of the luminosity  on the accretion column height for 
$\kappa_{\perp}=\kappa_{\rm T}$. The parameters used in the calculations are 
$M=1.4\,\msun$, $R=10^{6}$~cm, $\Lambda=0.5$, $l^{*}_{0}/l_0=1$, $\xi=1$, $f=1$. 
}
\label{pic:HofL}
\end{figure}

\subsection{Maximum luminosity}
\label{sec:Results}

Using the accretion column model described above we investigate the maximum accretion luminosity of magnetized NS as a function of the 
surface magnetic field strength. Higher accretion luminosity corresponds to a taller accretion column. 

The velocity profile inside the sinking region, which is defined by the parameter $\xi$, does not affect much the luminosity dependence on the accretion column height: its variations within the interval $1\leq\xi\leq 5$ lead to variations of accretion luminosity by a factor $\lesssim 2$ for a given column height. The higher the $\xi$, the lower the calculated luminosity value. If one takes scaled velocity structure obtained by \cite{BS1976} for the accretion column with constant scattering cross-section $\sigma=\sigma_{\rm T}$, the luminosity is smaller by a factor $<2$ than that given by $\xi=1$. Thus, we take $\xi=1$ in our calculations.

Theoretical dependences of X-ray pulsar luminosities on the accretion column height for various $B$-field strength are shown in 
Fig.\,\ref{pic:HofL}. All dependences start at relatively small accretion column heights corresponding to the critical luminosity $L^*$, which depends on the 
$B$-field strength as well \citep{2015MNRAS.447.1847M}. 
The lowest height is expected to be comparable with the thickness $d_0$ of the accretion channel near the NS surface.  
Below the critical luminosity, the energy is emitted by a hotspots at the NS surface. 
The luminosity dependences on the height of the accretion column are close to the approximate analytical dependence $L(H/R)$ given by 
equation~(\ref{eq:frh}) for the case of relatively low magnetic field strength ($B< 3\times 10^{12}$\,G). 
The deviations of the computed dependences from the analytical one can be explained by changes of accretion channel geometry, 
which is given by the ratio $l_0/d_0$ assumed to be constant in the approximate relation~(\ref{eq:frh}). 
The dependence of $l_0/d_0$ on the accretion rate varies according to the region where the accretion disc is interrupted by the NS magnetic field 
(see Section~\ref{sec:AccretionColumnBases}). If the accretion rate is relatively low and the disc is interrupted in the C-zone, 
the accretion column luminosity $L\propto f^{35/39}(H/R)$. 
However, if $\dot{M}$ becomes sufficiently high and the disc is disrupted in the A-zone, 
the luminosity grows with the column height much slower as $L \propto f^{7/16}(H/R)$. 
This transition from the C- and B-zones to the A-zone is clearly seen in Fig.~\ref{pic:HofL}. 
The transition luminosity depends on the magnetic field strength and is given by $L^{\rm tr}_{39}\approx 0.2\,B_{12}^{6/11}$.

The luminosity dependences on the accretion column height $L(H)$ for $B>10^{13}$~G are shifted to higher values 
because of a strong reduction of the effective electron scattering cross-section. 
The averaged scattering cross-section falls quickly at high $B$ and, therefore, the factor $\kappa_{\rm T}/\kappa_\perp$ grows substantially. 
We see, however, that the luminosity does not grow as much with the column height as it happens at lower $B$. 
This is related to the fact that with increasing $\dot{M}$ the optical thickness across the column grows leading to a 
higher inner temperature  and correspondingly higher scattering cross-section (because the photon peak becomes 
closer to the resonance) which leads to a reduction of the escaping flux.
Thus at some point the further growth of the column is not effective any more and luminosities higher than some maximum luminosity 
(for given $B$-field strength) cannot be supported by accretion column of any height. 

\begin{figure}
\centering 
\includegraphics[width=8.5cm]{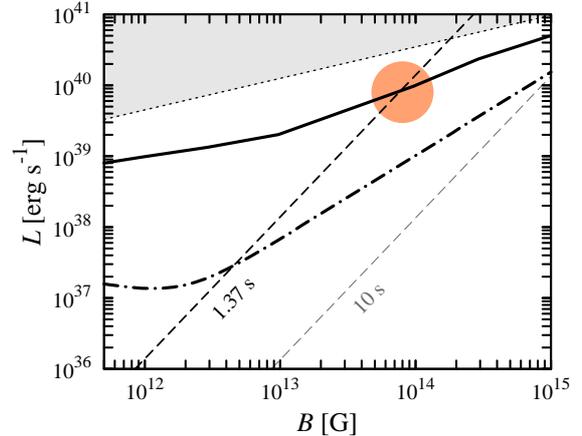} 
\caption{The maximum accretion luminosity $L^{**}$ as a function of the magnetic field strength is given by thick solid line.
In the interval $10^{13}<B<10^{15}\,{\rm G}$ it can be approximated as  $L_{39}^{**}\approx 0.35 B_{12}^{3/4}$. 
The lower limit on the X-ray luminosity  (\ref{eq:L_min}) shown by dashed lines 
is related to the inhibition of accretion by the propeller effect.
The upper limit (\ref{eq:Lmax_sph}) shown by the dotted line corresponds to the condition  
that the  spherization radius is smaller the magnetospheric radius. 
The critical luminosity $L^*$ when the accretion column starts to grow is shown
by the dashed-dotted line \citep{2015MNRAS.447.1847M}. 
The most likely position of the ULX in galaxy M82 \citep{Nature2014} is given by the orange circle, which corresponds to $B\simeq 8\times 10^{13}\,{\rm G}$, $H/R\simeq 1$, $l_0=4\times 10^5\,{\rm cm}$ and $d_0=1.7\times 10^4\,{\rm cm}$. 
Parameters: $M=1.4 \msun$, $R=10^{6}$~cm, $\Lambda=0.5$, $l^{*}_{0}/l_0=1$, $\xi=1$, $f=1$.}
\label{pic:final}
\end{figure}

The dependence of the maximum luminosity on the surface magnetic field is shown in Fig.~\ref{pic:final} 
by a thick solid line.  
We limited the maximal accretion luminosity for low $B$-field strength by hand at $H/R = 1$, because of a much slower 
luminosity grow at  $H/R>1$ and a questionable validity of our model for such high columns. 
However, in any case we see that the maximum column luminosity $L_{39}^{**}\sim 0.35 B_{12}^{3/4}$ 
can easily reach value in excess of $10^{40}\,$\ergs\ for magnetar-like magnetic fields. 
The parameters of the model such as the exact velocity profile in the sinking region of the accretion column and the 
fraction of photon in each polarization state do not affect much the final results.
The relation between the luminosity and the $B$-field can be reversed implying 
that for a given luminosity ($L_{39}\gtrsim 2$) there is a lower limit on the magnetic field: $B_{12}\gtrsim 4\,L^{4/3}_{39}$.

\section{Discussion}
\label{sec:Disc}

\subsection{An admissible accretion luminosity range}
\label{sec:Range}

We have shown in the previous section that a tall accretion column can radiate in excess of $10^{40}\,$\ergs\ for the 
magnetar like magnetic fields $B\gtrsim10^{14}$~G. 
However, we have not addressed the question if a necessary high accretion rate 
is consistent with the presence of the column. The problem is that  at high accretion rate the 
accretion disc may become locally super-Eddington outside of the magnetosphere
producing strong outflows \citep{SS1973,2007MNRAS.377.1187P}  
and reducing the NS luminosity. 
Additional problem arises because in this case the disc becomes very thick at $R_{\rm m}$ leading 
to a nearly spherical accretion, which is  limited by the Eddington luminosity. 
The correction of the opacity at high $B$ will not play any role here just because the magnetic field 
drops rapidly  with the distance and at some radius the cyclotron resonance will appear in the X-ray range 
where most of the photons are escaping.

The condition for the absence of the outflows outside the magnetosphere can be written as 
\be\label{eq:Rsp_Rm}
R_{\rm sp} < R_{\rm m}, 
\ee
where $R_{\rm sp}$ is the spherization radius where the radiation force  is still balanced by gravitation.
Its simplest estimate is based on equating the scale-height of the radiation-dominated 
disc to the radial distance \citep{SS1973}: 
\be
R_{\rm sp}= \frac{3}{8\pi}\frac{\kappa_{\rm T}}{c}\dot{M}  =7.5 \times 10^{6}\frac{L_{39}R_{6}}{m}\, \mbox{cm}.
\ee 
It may be smaller by a factor of a few if one accounts for the energy advection \citep{2007MNRAS.377.1187P}. 
Using the condition (\ref{eq:Rsp_Rm}) we get an approximate upper limit for the accretion luminosity:
\be \label{eq:Lmax_sph}
L^{\max}_{39}\simeq 5.7\,\Lambda^{7/9}\, B^{4/9}_{12}m^{8/9} R^{1/3}_{6} 
\ee
shown by a dotted line in Fig.~\ref{pic:final}. 
We see that the theoretical maximum column luminosity is always below the limit coming 
from this spherization constraint and therefore our model is self-consistent. 

In addition to the upper limit on the luminosity, there should exist a lower limit too. 
At low accretion rates, the spin frequency of the NS may exceed the Keplerian frequency at the magnetospheric radius. 
Then the accretion of matter may be inhibited by the centrifugal force exerted by the rotating magnetosphere and 
the matter then may be expelled from the system  due to the ``propeller effect'' \citep{IS1975}.
This dramatically reduces the accretion luminosity. 
The propeller starts to operate when the magnetospheric radius $R_{\rm m}$ becomes larger than 
the corotation radius
\be
R_{\rm C}=\left(\frac{GMP^2}{4\pi^2}\right)^{1/3}\simeq 1.5\times 10^{8}\,m^{1/3} P^{2/3}\,\mbox{cm},
\ee
where $P$ is a NS spin period in seconds. 
This happens at luminosity 
\be\label{eq:L_min}
L^{\min}_{39} \simeq 7\times 10^{-2}\,\Lambda^{7/2}m^{-2/3}R_6^5 B^{2}_{12}P^{-7/3} , 
\ee
which defines the likely lower limit on X-ray luminosity of a binary system. 
 
During the accretion process a NS in a binary system is spun up to the period determined by the condition $R_{\rm m}\simeq R_{\rm C}$,
when the accretion luminosity  is given by equation (\ref{eq:L_min}). 
For the Eddington accretion rate relaxation happens very rapidly, within 
$\tau_{\rm rel}\approx 300\, I_{45}m^{1/3}\mu_{30}^{-4/3}$~yr, where $I_{45}=I/10^{45}$~g\,cm$^2$ is the NS moment of inertia  
and $\mu_{30}=\mu/10^{30}$~G\,cm$^3$ is the NS dipole magnetic moment \citep{1982SvA....26...54L}. 
Of course, the luminosity may exceed the minimum value  (\ref{eq:L_min}) 
over short time intervals when the accretion rate increases. 
 
We conclude that the X-ray pulsar luminosity is likely to be bound from below by $L^{\min}$ 
and from above by $L^{**}\approx 0.35 B_{12}^{3/4}$.  
It is clear from Fig.~\ref{pic:final} that the stronger the magnetic field, 
the narrower the possible luminosity range for a given NS spin period. 
A likely maximum accretion luminosity given by the interception of the two constraints is then 
\be 
L_{39} \approx 1.6\, m^{2/5} P^{5/4} R_6^{-3} \Lambda^{-21/10},  
\ee
corresponding to the magnetic field  
\be 
B_{12}  \approx 7.5\, m^{8/15} P^{28/15} R_6^{-4} \Lambda^{-14/5} . 
\ee 
Thus the luminosities of X-ray pulsars in excess of $10^{40}\,$\ergs\ would definitely require 
an extremely strong $B$-field ($B>10^{14}$~G) and a large spin period ($P>1$~s).

\subsection{X-ray pulsars and ULXs}

All known X-ray pulsars can roughly be divided into two big groups: persistent and transient sources. 
Among the  brightest persistent sources we mention here Cen X-3, SMC X-1 and LMC X-4. 
The only source in this group with the known magnetic field is Cen X-3, which shows the cyclotron line around 28 keV \citep{1992ApJ...396..147N} 
that corresponds to the corrected for the redshift magnetic field $B\sim 3.2\times10^{12}$~G. 
The brightest X-ray pulsars in the Magellanic Clouds do not show any cyclotron features in their spectra and 
hence the magnetic fields there are unknown. 
The maximal observed luminosities in this group range from $\sim5\times10^{37}\,$\ergs\ for 
Cen X-3 to $\sim10^{39}\,$\ergs\ for SMC X-1 and LMC X-4 during outbursts.

Another group of transient sources mainly consists of NS in binary systems with the Be companions. 
They are known to exhibit rare non-periodic type II outbursts with the peak X-ray luminosity reaching the 
Eddington value \citep[for a review see][]{2011Ap&SS.332....1R}. 
The brightest sources in this group are 4U 0115+63 and V 0332+53 with the peak luminosities as high as a few $\times10^{38}\,$\ergs\ 
and a clear evidence of the high accretion column 
\citep[see e.g.,][]{2006MNRAS.371...19T, 2007AstL...33..368T, 2010MNRAS.401.1628T, Poutanen2013,2015MNRAS.452.1601P}. 
The magnetic fields are known from the cyclotron lines: $\sim 1.3\times10^{12}$ G for 4U 0115+63 \citep{1983ApJ...270..711W} and 
$\sim 3.2\times10^{12}$ G for V 0332+53 \citep{1990ApJ...365L..59M}.

Apart of these two groups of X-ray pulsars we can mention a unique transient source GRO J1744--28 showing both X-ray pulsations and 
unusual short bursts. 
The broadband observations of this source revealed a cyclotron absorption line in the energy spectrum at 
$E \approx 4.5$~keV \citep{2015MNRAS.449.4288D, 2015arXiv150309020D}, which corresponds to 
$B\simeq 5\times 10^{11}$~G. 
The maximum luminosity during the outbursts was $\sim (3\div 4)\times 10^{38}\,$\ergs. 
The observed luminosities of all these pulsars are in good agreement with our limits. 
 
It was shown above that the accreting NS may have accretion columns radiating in excess of $10^{40}\,$\ergs\ 
for magnetar-like magnetic fields  $B_0\gtrsim 10^{14}$~G (see Fig.~\ref{pic:final}). 
In other words, some ULXs can actually be extreme X-ray pulsars as indeed was recently 
observed for X-2 source in galaxy M82  \citep{Nature2014}. 
Its high pulsating luminosity $\sim 4.9\times 10^{39}\,$\ergs\  implies according to our model 
the magnetic field $B\gtrsim 10^{14}$~G. 
The observed pulsation period of $1.37$~s makes the magnetic field of $\sim 10^{14}$~G the only possible option for this particular ULX, 
because the higher field strength leads to the propeller effect (see Section~\ref{sec:Range}). 
Because the ULX is very likely to be close to the equilibrium state \citep{Nature2014,1982SvA....26...54L}, 
 an estimation of its magnetic field based on the analysis of the period $P$ and the period derivative $\dot{P}$ 
is not valid since one need to know exact physics of disc and magnetosphere interaction and the estimations of magnetic field strength become strongly model dependent near the equilibrium \citep{2015arXiv150708627P,2015MNRAS.449.2144D,2015MNRAS.448L..40E}. Transition between accretion and propeller regime has been detected recently in M82 X-2 \citep{2015arXiv150708288T} confirming our estimation of the magnetic field strength of $\sim 10^{14}\,{\rm G}$.
A high accretion rate needed to power high ULX luminosities ($L\sim 10^{39}-10^{41}\,$\ergs) from pulsars 
imply a large thickness of the accretion disc on its inner edge. 
This results in reprocessing of a significant part of pulsar radiation in the disc, which might lead to spectral deviations  
comparing to the intrinsic spectrum of the accretion column.

\section{Summary}

We have constructed a simplified model of the accretion column on a magnetized NS. 
We based the model on previous suggestions by \citet{BS1976} and \citet{LS1988} 
that the column is vertically supported by the Eddington flux. 
We use diffusion approximation to evaluate the radiation flux through the sides of the column.  
For the first time, we use the exact Compton scattering cross-sections in strong magnetic field and 
find the column temperature structure and the cross-sections self-consistently via an iterative procedure. 
As a result we have obtained the accretion column luminosity as a function of the magnetic field strength and the column height. 
We also evaluate for a given surface magnetic field the maximum possible column luminosity that is reached 
when the column height becomes comparable to the NS radius. 
We showed that the column may radiative in excess of $10^{40}\,$\ergs\ in the case of the magnetar-line 
magnetic field  $\gtrsim 10^{14}$~G. 

We also evaluate the possible range of accretion luminosity  in X-ray pulsars. 
It is limited from above by the condition that the accretion disc does not spherize outside the magnetosphere. 
This condition, however, gives the limiting luminosity above our upper limit  thus confirming the consistency of our calculations. 
The pulsar luminosity may also be limited from below by the propeller effect. 

The range of possible pulsar luminosities becomes narrower for stronger magnetic fields. 
The surface magnetic field strength can be limited from below if one knows the luminosity of the object: $B_{12}\gtrsim 4\,L^{4/3}_{39}$.
Thus the X-ray pulsar luminosity exceeding a few $10^{39}\,$\ergs\ requires magnetic field $\gtrsim 10^{13}$~G,
while the ULX luminosities of $10^{40}\,$\ergs\ would be consistent only with the magnetar-like fields  $\gtrsim 10^{14}$~G.
On the other hand, the magnetic field cannot be too large, because of the propeller effect. 
We showed that the estimated X-ray luminosity of ULX X-2 in galaxy M82 \citep{Nature2014} of $\approx 10^{40}\,$\ergs\  
can be consistent with our model only for $B\approx 10^{14}$~G.  

Because of the set of constraints it is likely that the luminosity $L\simeq (1\div 3)\times 10^{40}\,$\ergs\ 
is a good guess for the NS maximum possible accretion  luminosity. 
Interestingly, this value coincides with the cut-off observed in the  X-ray luminosity function of high-mass X-ray binaries, 
which does not show any features at lower luminosities \citep{Mineo2012}. 
Therefore, it is natural to suggest that accreting NSs are not rare among ULXs 
and their fraction  is probably close to their fraction among the high-mass X-ray binaries.

\section*{Acknowledgements}

This research was supported by the Magnus Ehrnrooth Foundation (VFS), 
the Jenny and Antti Wihuri Foundation (VFS), 
the Russian Science Foundation grant 14-12-01287 (AAM, SST), 
the Academy of Finland grant 268740 (JP), and 
the German research Foundation (DFG) grant WE 1312/48-1 (VFS). 
Partial support comes from the EU COST Action MP1304 ``NewCompStar''.




\label{lastpage}

\end{document}